\documentclass[aps, pra, twocolumn, showkeys, 10pt]{revtex4-2}

\usepackage{amssymb}
\usepackage{xspace}
\usepackage{CJKutf8}
\usepackage[utf8]{inputenc}
\usepackage{lineno}
\usepackage{url}
\usepackage{graphicx}
\usepackage{xcolor}
\usepackage{hyperref}

\newcommand{\model}{AstroSage-Llama-3.1-8B\xspace}
\newcommand{\qa}{Q$\&$A\xspace}
\newcommand{\score}{$80.9\%$\xspace}
\newcommand{\refresp}[1]{\textcolor{black}{#1}}

\begin{document}

\begin{CJK}{UTF8}{gbsn}

\title{Achieving GPT-4o Level Performance in Astronomy with a Specialized 8B-Parameter Large Language Model}

\author{Tijmen de Haan$^{1,2,*}$, 
Yuan-Sen Ting$^{3,4}$,  
Tirthankar Ghosal$^{5}$, 
Tuan Dung Nguyen$^{6}$, 
Alberto Accomazzi$^{7}$,
Azton Wells$^{8}$,
Nesar Ramachandra$^{9}$, 
Rui Pan$^{9}$, 
Zechang Sun$^{10}$
}

\affiliation{$^{1}$Institute of Particle and Nuclear Studies (IPNS), High Energy Accelerator Research Organization (KEK), Tsukuba, Ibaraki, Japan}
\affiliation{$^{2}$International Center for Quantum-field Measurement Systems for Studies of the Universe and Particles (QUP-WPI), High Energy Accelerator Research Organization (KEK), Tsukuba, Ibaraki, Japan}
\affiliation{$^{3}$Department of Astronomy, The Ohio State University, Columbus, OH, USA}
\affiliation{$^{4}$Center for Cosmology and AstroParticle Physics (CCAPP), The Ohio State University, Columbus, OH, USA}
\affiliation{$^{5}$National Center for Computational Sciences, Oak Ridge National Laboratory, Oak Ridge, TN, USA}
\affiliation{$^{6}$Department of Computer and Information Science, University of Pennsylvania, Philadelphia, PA, USA}
\affiliation{$^{7}$Center for Astrophysics, Harvard \& Smithsonian, Cambridge, MA, USA}
\affiliation{$^{8}$Computational Science Division, Argonne National Laboratory, Lemont, IL, USA}
\affiliation{$^{9}$Department of Computer Science and Engineering, Hong Kong University of Science and Technology, Kowloon, Hong Kong}
\affiliation{$^{10}$Department of Astronomy, Tsinghua University, Beijing, People's Republic of China}
\affiliation{$^{*}$Corresponding author: \href{mailto:tijmen.dehaan@gmail.com}{\normalfont tijmen.dehaan@gmail.com}}
           
\begin{abstract}
\model is a domain-specialized natural-language AI assistant tailored for research in astronomy, astrophysics, cosmology, and astronomical instrumentation. Trained on the complete collection of astronomy-related arXiv papers from 2007-2024 along with millions of synthetically-generated question-answer pairs and other astronomical literature, \model demonstrates remarkable proficiency on a wide range of questions. \model scores \score on the AstroMLab-1 benchmark, greatly outperforming all models---proprietary and open-weight---in the 8-billion parameter class, and performing on par with GPT-4o. This achievement demonstrates the potential of domain specialization in AI, suggesting that focused training can yield capabilities exceeding those of much larger, general-purpose models. \model is freely available, enabling widespread access to advanced AI capabilities for astronomical education and research. 
\end{abstract}


\maketitle

\end{CJK}

\section{Introduction}
\label{introduction}

Large-language model (LLM) assistants are rapidly gaining traction across all sectors of knowledge work worldwide. In astronomy, these models are used for providing factual information, as programming assistants, for brainstorming ideas, and for providing explanations tailored to the level of understanding or preferred style of the user. LLMs exhibit a remarkable robustness, often delivering useful outputs even when the input is malformed, lacks context, or contains inaccuracies.

Despite their potential, the development of specialized LLMs has been limited due to their recent emergence and the substantial resources required for training. Previous studies \citep{fuTinyTitansCan2024,hoffmannTrainingComputeOptimalLarge2022,schickItsNotJust2021,turcWellReadStudentsLearn2019} have shown that models narrowly tailored to a specific domain can perform on par with, or even exceed, much larger general-purpose models. This suggests that a large, highly domain-specific model could achieve state-of-the-art performance.

In astronomy, however, high-performing specialized language models have not yet been achieved. While models like AstroLLaMA \citep{nguyenAstroLLaMASpecializedFoundation2023,perkowski_astrollama-chat_2024} have gained attention, they lack comprehensive benchmarking of their astronomical knowledge recall capabilities. Recent studies \citep{pan_astromlab_2024} have shown that many of these models, due to limited specialized training data and fine-tuning for instruction-following, suffer from either catastrophic forgetting or an inability to follow precise question-answering instructions, often performing worse than their baseline models (in this case, the Llama models).

Building on the previous efforts of \textsc{cosmosage} \citep{de_haan_cosmosage_2025} and AstroLLaMA, we have developed \model, a natural language assistant specialized in astronomy, astrophysics, cosmology, and astronomical instrumentation. For the remainder of this paper, we will refer to these subdomains collectively as ``astronomy''. Through the use of a substantially more extensive and well-curated training dataset, we demonstrate for the first time that our specialized language model significantly outperforms baseline models in downstream tasks.

In the long term, we envision an agentic research assistant capable of autonomously conducting literature reviews, identifying relevant hypotheses, carrying out data analysis, and even formulating new research questions. The development of such scientific agents (LLMs capable of solving scientific problems end-to-end) is already a rapidly growing field in astronomy. Recent studies have shown promising results in automating research tasks, such as analyzing James Webb Space Telescope data through multi-agent collaboration and self-play reinforcement learning \citep{sun2024}. However, these studies have been largely constrained by \refresp{the substantial costs associated with utilizing proprietary models at high volume}.

Realizing this level of agency will require extensive experimentation and careful optimization. Given the substantial compute costs and data requirements inherent in large-scale model training, keeping the model size manageable while maintaining high performance is crucial. Our approach, demonstrated through astronomy knowledge recall, shows that specialized models can achieve state-of-the-art performance in specific domains. This not only makes the development of advanced research assistants more feasible but also ensures their accessibility to a wider range of institutions and researchers, potentially transforming the landscape of astronomical research and education.

\section{Continued Pretraining}
\label{sec:pretraining}

For \model, we selected Meta's Llama-3.1 8-billion parameter model \citep{dubeyLlama3Herd2024b} as our foundation model. This base model was chosen for both its strong general-purpose capabilities and its availability under the permissive Llama 3.1 Community License. Furthermore, among models in the 8-billion parameter class, it demonstrated superior performance in astronomical knowledge recall compared to both general-purpose models \citep{tingAstroMLab1Who2024} and specialized astronomical LLMs \citep{pan_astromlab_2024}, making it an ideal baseline. 

To begin the development process, we first focused on curating, obtaining, and cleaning a continued pretraining (CPT) dataset.

\subsection{Dataset Preparation}
\label{sec:dataset}

The scaling laws of \citet{hoffmannTrainingComputeOptimalLarge2022} show that model capability increases predictably with increased training data volume and computational resources. More recently, it has been found \citep{raeScalingLanguageModels2021,brown_language_2020,penedoRefinedWebDatasetFalcon2023, liScalingFilterAssessingData2024} that the power-law index of these scaling laws depends on data quality. Therefore, our general approach to assembling a corpus was to focus on maintaining a high quality threshold, maximizing data volume at that quality level. 

We employed a multi-faceted approach to create a comprehensive, high-quality, high-variety CPT corpus. The primary components of our dataset include:

\begin{itemize}
    \item Approximately 250,000 arXiv preprints from 2007-2024 with primary or cross-listed categories in astro-ph (astrophysics) or gr-qc (general relativity and quantum cosmology). \refresp{For preprints with multiple versions, we selected the most recent.} We deliberately excluded the Annual Review of Astronomy and Astrophysics papers used in \citet[AstroMLab-1]{tingAstroMLab1Who2024} to generate the benchmark questions, ensuring our evaluation would test the model's ability to generalize knowledge rather than recall specific source materials. 
    \item \refresp{Relevant articles from a depth-2 search through Wikipedia's astronomy and astrophysics categories. We include articles that directly belong to these categories or their immediate subcategories, then further add linked articles. We perform keyword-based exclusion on a few tens of custom keywords such as ``deity'', ``cyberspace'', ``mythological'', ``sports'' and so on, yielding a total of nearly 30,000 articles.}
    \item A selection \refresp{of $\sim$800} textbooks that are available as PDFs or ebooks online.
\end{itemize}

For the vast majority of the dataset, the rendered PDF files were converted to \refresp{Mathpix Markdown} using Nougat OCR \citep{blecherNougatNeuralOptical2023}. For the remaining sources, the data was either already in Markdown format, or was left as plain text.

\subsection{Pretraining run}
\label{sec:cpt_run}
 
The pretraining was conducted on the ORNL OLCF Frontier 1.6 exaflop supercomputer, leveraging its substantial computational resources. We used 184 nodes simultaneously, each of which is equipped with four AMD MI250X, which in turn have 2 Graphics Compute Dies (GCDs) each for a total of 8 GCDs per node, or a total of 1472 GCDs.

Further statistics and our choices of pretraining hyperparameters are summarized in Table~\ref{tab:hyper}.

\begin{table}[ht]
\centering
\begin{tabular}{|l|c|}
\hline
\textbf{Hyperparameter}          & \textbf{Value}            \\ \hline
Sequence length                  & 8,192 tokens              \\ \hline
Micro batch size                 & 3                         \\ \hline
Epochs                           & 2                         \\ \hline
Learning rate                    & 1.5e-4                    \\ \hline 
Base model                       & Llama 3.1 8B              \\ \hline
Optimizer                        & AdamW                     \\ \hline
Adam beta2                       & 0.95                      \\ \hline
Adam epsilon                     & 1e-5                      \\ \hline
Learning rate schedule                     & Constant with \\ & quadratic warmup   \\ \hline
Max gradient norm                & 3.0                       \\ \hline
Weight decay                     & 0.001                     \\ \hline
Warmup steps                     & 40                        \\ \hline
Precision                        & BF16                \\ \hline
FSDP                             & Full shard, auto wrap     \\ \hline
\hline
\textbf{Resource}                & \textbf{Value}            \\ \hline
Plaintext filesize               & 19.9 GB                   \\ \hline
Token count                      & 3.3 billion               \\ \hline
Nodes                            & 184                       \\ \hline
GCDs (effective \# of GPUs)  & 1,472                     \\ \hline
Training wall time               & 10 hours                  \\ \hline
Total time spent                 & 11.5 hours                \\ \hline
Effective GPU-hours spent        & 14,872 hours              \\ \hline
VRAM Usage                       & 96\% of 64 GB/GCD         \\ \hline
\end{tabular}
\caption{Summary of pretraining hyperparameters and resource usage.}
\label{tab:hyper}
\end{table}

The Llama-3.1 tokenizer, which uses a variant of tiktoken \citep{noauthor_openaitiktoken_2025} with UTF-flexible encoding, was sufficient for our purposes, so we did not introduce any astronomy-specific tokens to the vocabulary. We considered incorporating arXiv identifiers (in the format arXiv:YYMM.numbervV) for papers in the training set but ultimately decided against it due to the substantial increase in vocabulary size this would entail. Future work may explore expanding the vocabulary to include numerical representations relevant to astronomy, such as common quantities and units.

During training, we tracked the loss function and step sizes. Minimal hyperparameter tuning was performed, relying on the hyperparameters from \citet{de_haan_cosmosage_2025} for all parameters other than the learning rate schedule. \refresp{The warmup period of 40~samples was determined empirically from the loss curve of an earlier, canceled CPT run. The peak} learning rate was extrapolated from smaller runs, but remained problematic, as Frontier requires high levels of parallelization with short wall times. This caused our initial runs to suffer from either insufficient learning due to a low \refresp{peak} learning rate, or catastrophic exploding gradients due to an excessively high \refresp{peak} learning rate. The final run described in this work used a tuned learning schedule with a \refresp{peak} learning rate as high as possible but still allowing convergence. In future efforts, we plan to further optimize the efficiency of the training procedure along the lines of the work presented in \citet{dashOptimizingDistributedTraining2023}, incorporating tensor, pipeline, and data paralellism through libraries such as Megatron-Deepspeed \citep{noauthor_deepspeedaimegatron-deepspeed_2025, smith_using_2022}. We also aim to request a longer walltime with lower parallelization factor in order to be able to reduce loss and improve downstream performance. 

\subsection{Dataset Cleaning}

We followed the cleaning procedures from \citet{de_haan_cosmosage_2025}, including a perplexity-based cleaning approach. This method first splits the corpus into individual paragraphs and calculates their respective perplexity scores. Perplexity measures how well a language model can predict a given text sequence. Lower perplexity indicates text that follows expected patterns of natural language, while very high perplexity values often signal anomalous or corrupted content. Outliers with such high perplexity frequently stem from OCR errors, malformed text, or non-prose content such as tables.

\begin{figure}
    \centering
    \includegraphics[width=0.45\textwidth]{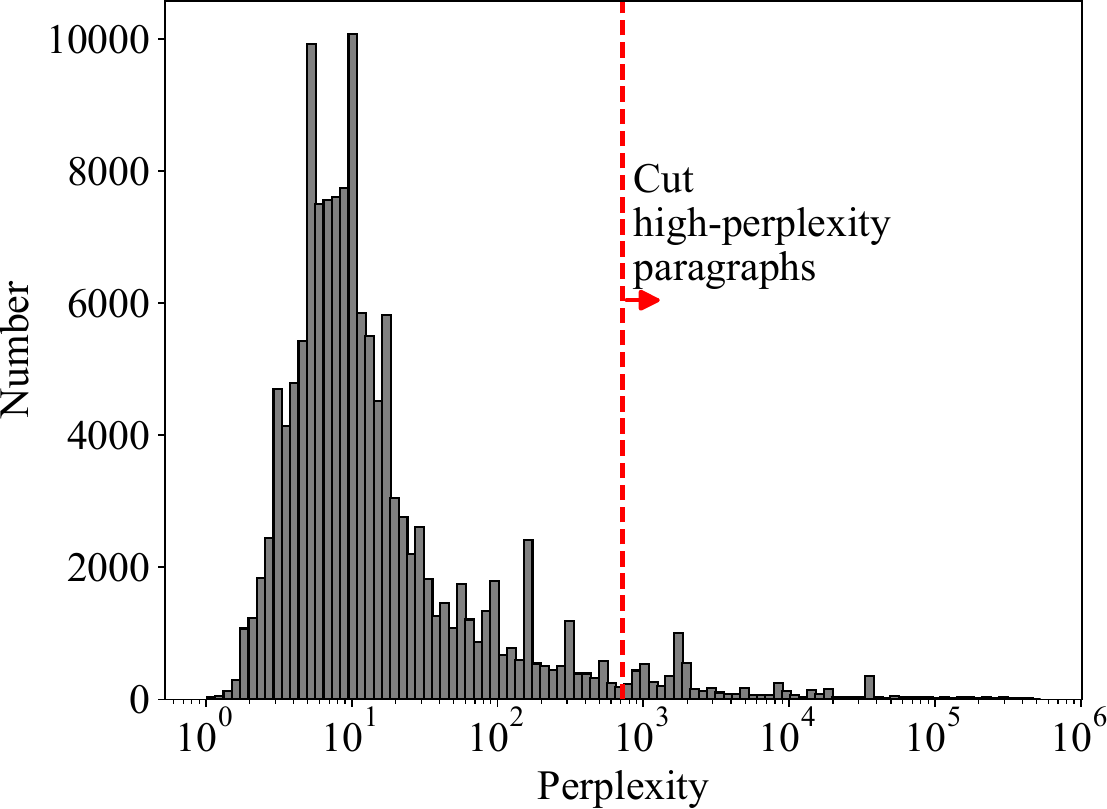}
    \caption{Histogram of paragraph-level perplexity used for dataset cleaning. The perplexity is calculated as $\exp(-\langle\ln P\rangle)$, where $\langle\ln P\rangle$ is the average log probability per token in each paragraph. The red dashed line indicates the manually chosen threshold used to filter out high-perplexity paragraphs, which comprised approximately 2\% of the total data volume. Paragraphs with perplexity above this threshold were removed from the dataset. Note that the highest observed perplexity values (around $10^{12}$) extend beyond the right edge of the plot.}
    \label{fig:ppx}
\end{figure}

Based on the distribution of perplexity scores (Figure~\ref{fig:ppx}), \refresp{as well as manual inspection of some examples,} we established a threshold that excluded the top 2\% of paragraphs with the highest perplexity scores. We then reconstructed each document using only the paragraphs below this threshold. This cleaning procedure removed approximately 2\% of the total data volume. Figure~\ref{fig:ppx} illustrates this process by showing the distribution of perplexity scores and our chosen threshold.

\section{Supervised Fine-tuning}
\label{sec:sft}

To improve the model's ability to follow instructions and answer questions effectively, we performed supervised fine-tuning (SFT). In this process, the model was trained to predict appropriate responses to given prompts, learning from a collection of high-quality example conversations. Below, we describe our approach to generating and curating the SFT dataset.

\subsection{SFT Dataset}

\citet{pan_astromlab_2024} identified a critical limitation in the AstroLLaMA series of specialized astronomical LLMs: their inability to outperform even their own starting base models, partly due to inadequate SFT. While these models showed marginal improvements in basic next-token prediction tasks, they performed worse than their baseline models on instructional question-and-answer (\qa) tasks, even for straightforward astronomical knowledge recall. This shortcoming fundamentally undermines the purpose of specialized training. Therefore, in our study, we paid particular attention to the SFT process, generating training datasets orders of magnitude larger than previously available in astronomy.

The largest component of our SFT dataset consists of \qa pairs. Using the method from \citet{de_haan_cosmosage_2025}, we generated over 11 million synthetic \qa pairs from papers in our CPT dataset. These \qa pairs were then evaluated using an LLM based on four criteria:
\begin{enumerate}
    \item Correctness: The factual accuracy of the answer in relation to the question, ensuring that each \qa pair adheres to current scientific understanding and accurately reflects the information presented in the source material.
    \item Stand-alone: The ability of each \qa pair to be understood in isolation, without needing additional context beyond the content provided. This ensures that an \refresp{astronomy expert} could answer the question based solely on the information within the pair.
    \item Pertinence: The relevance and importance of the question to researchers or students \refresp{in astronomy.} We want the questions to be ones that a professional in the field might ask or find valuable for deeper understanding.
    \item Overall Quality: a holistic assessment of the \qa pair. This criterion is intended to capture aspects that may not be explicitly covered by correctness, stand-alone completeness, or pertinence but contribute to the overall usefulness of the \qa pair.
\end{enumerate}
The LLM was presented with several hand-written examples of ratings in the form Question--Answer--Score--Explanation, followed by the Question--Answer pair to be judged. The model was then asked to complete the score values. A small number of resulting scores were verified and confirmed to be sufficiently accurate. Only \qa pairs with perfect scores in all four categories were kept, resulting in 8.8 million high-quality \qa pairs. \refresp{Below are a few examples.}

\begin{itemize}
    \item \textbf{Q:} \refresp{What are the three distinct phases of galaxy formation identified in the EAGLE simulation, and how are they primarily determined?}
    
    \textbf{A:} \refresp{The three distinct phases of galaxy formation identified are:}
    \begin{enumerate}
        \item \refresp{Disorganized growth in low-mass galaxies (stellar mass less than approximately $10^{9.5} M_\odot$), characterized by random stellar motions and dominated by in-situ star formation and mergers; }
        \item \refresp{Evolution towards a disc-dominated morphology in intermediate-mass galaxies (stellar mass between $10^{9.5} M_\odot$ and $10^{10.5} M_\odot$), with bulge formation associated with mergers; }
        \item \refresp{High-mass galaxies (stellar mass greater than approximately $10^{10.5} M_\odot$) experience a slowdown in in-situ star formation, transitioning to spheroidal morphologies driven by the accretion of ex-situ stars. These phases are primarily determined by stellar mass rather than redshift.}
    \end{enumerate}

    \item \textbf{Q:} \refresp{What are the expected contributions of AGN feedback to the heating and dynamics of the hot ISM?}
    
    \textbf{A:} \refresp{AGN feedback is anticipated to play a significant role in regulating the cooling of the hot ISM, with energy released during accretion often exceeding starburst-driven heating and driving turbulence and outflows that influence the dynamics of the ISM.}

    \item \textbf{Q:} \refresp{What are the applications of the Augmented Lagrangian Perturbation Theory (ALPT) in cosmology?}
    
    \textbf{A:} \refresp{ALPT can be used for setting initial conditions for N-body simulations, generating mock galaxy catalogs, analyzing cosmic structures, and reconstructing primordial density fluctuations for baryon acoustic oscillation studies.}

    \item \textbf{Q:} \refresp{What are the three effects that anisotropies in the optical depth produce in the CMB?}
    
    \textbf{A:} \refresp{Anisotropies in the optical depth produce three effects in the CMB: (i) screening of the temperature and polarization fluctuations observed today by an overall factor of $\exp(-\tau(\hat{n}))$; (ii) Thomson scattering, producing new E-mode polarization from the local temperature quadrupole, with subdominant contributions to B-modes from reionization effects; and (iii) new temperature anisotropy generated from the radial motion of ionized bubbles relative to the observer.}

\end{itemize}

We also included a filtered version of Infinity-Instruct-7M \citep{beijingacademyofartificialintelligencebaaiBAAIInfinityInstructDatasets2024}, keeping only entries with at least 70\% alphanumeric characters, as well as filtering out entries with certain keywords. The inclusion of this dataset was to ensure that \model would gain instruction-following abilities such as multi-turn conversation.

Additionally, we generated synthetic summaries for all of the papers in the CPT dataset. The SFT dataset for these summaries consists of a user prompt asking to summarize a certain paper, with the assistant completion being a small preamble followed by the summary. The user prompt was created through a series of random choices about the way the question is asked, a small preamble, and the way in which the paper is referenced, yielding a high variety of question styles. 

Furthermore, we generated a metadata-based dataset, where again a series of custom rules and random selections result in diverse questions about titles, dates of publication, arXiv IDs, and first author names from the papers in the CPT dataset. This was included in an effort to memorize the paper metadata so that users can reference papers in their conversations with \model. \refresp{However, we find that after training, the model is unable to reliably answer metadata-based questions such as producing a first author name from an arXiv ID. This failure to memorize could be ameliorated in future training runs by increasing the learning rate or number of epochs.}

These datasets were combined with five further datasets which were assembled by hand from various sources on the web. The combined dataset comprised approximately 2 billion tokens.

\subsection{SFT procedure}

The SFT process was also conducted on the Frontier supercomputer with a configuration summarized in Table~\ref{tab:hyper2}.

\begin{table}[ht]
\centering
\begin{tabular}{|l|c|}
\hline
\textbf{Hyperparameter}          & \textbf{Value}            \\ \hline
Epochs                           & 6                         \\ \hline
Learning rate                    & 1e-4                      \\ \hline
Base model                       & CPT model    \\ \hline
Learning rate schedule                      & Cosine with \\ & quadratic warmup   \\ \hline
Weight decay                & 0.0                       \\ \hline
\hline
\textbf{Resource}                & \textbf{Value}            \\ \hline
Plaintext filesize               & 9.8 GB                    \\ \hline
Token count                      & 2.0 billion               \\ \hline
Training wall time               & 9.5 hours                 \\ \hline
Total time spent                 & 11.5 hours                \\ \hline
Effective GPU-hours spent        & 13,738 hours              \\ \hline
\end{tabular}
\caption{Summary of supervised fine-tuning hyperparameters and resource usage. Parameters that are not stated here were kept the same as in Table~\ref{tab:hyper}.}
\label{tab:hyper2}
\end{table}

\begin{figure*}
    \centering
    \includegraphics[width=\textwidth]{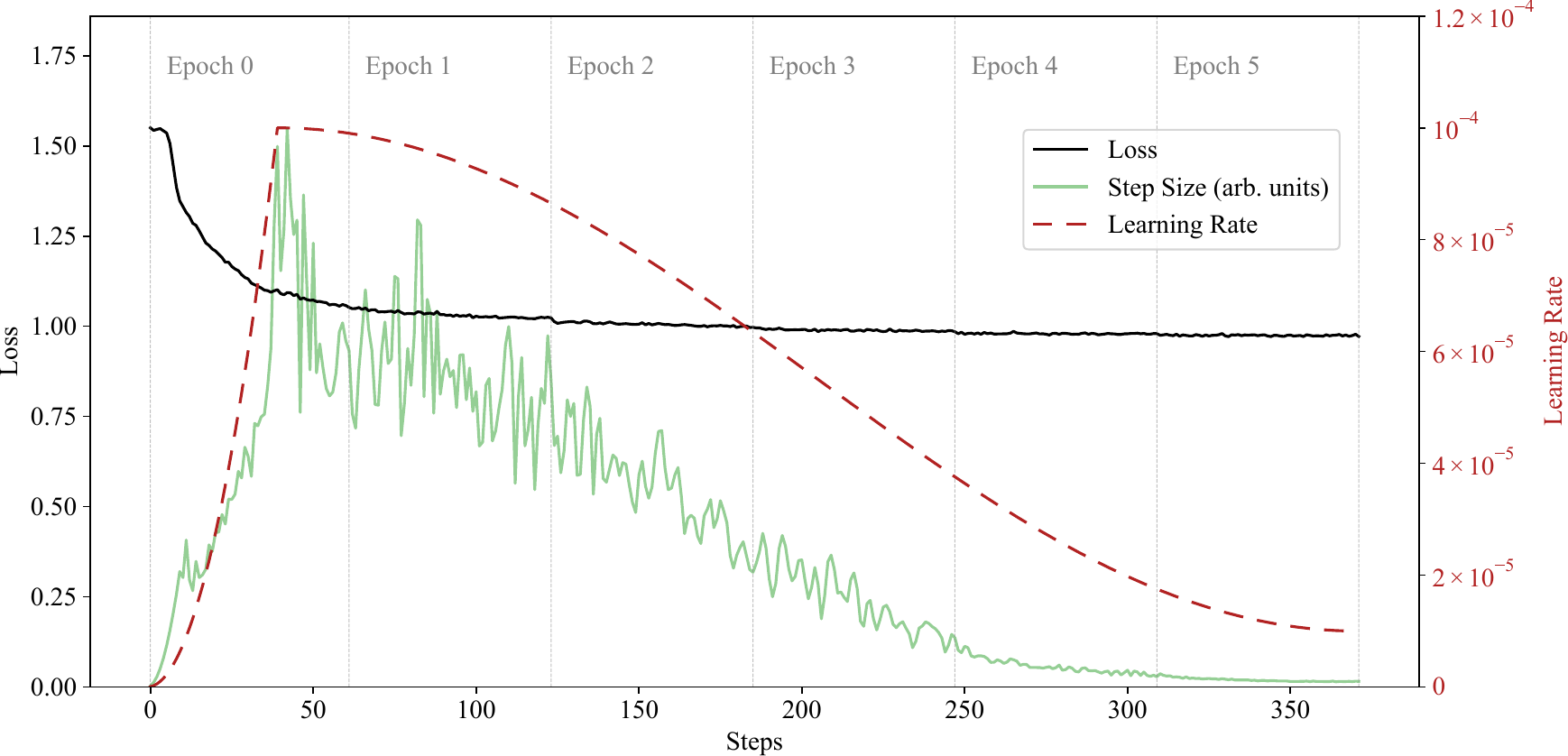}	
    \caption{Supervised fine-tuning loss curve. The learning rate schedule, shown as the dashed brick red curve, follows a quadratic warmup followed by a cosine schedule that ends at 10\% of the peak learning rate. The peak learning rate was chosen to prevent exploding gradients. The loss curve in black shows no significant signs of overfitting, as evidenced by minimal discontinuities at epoch boundaries. The green curve represents the L2 norm of the parameter update, shown in arbitrary units.}
    \label{fig:curve_sft}
\end{figure*}

Throughout the fine-tuning process, we again monitored loss and step sizes, which are shown in Figure~\ref{fig:curve_sft} (CPT curves look similar). The learning was---like in \S\ref{sec:cpt_run}---limited by the maximum wall-time allowed by the Frontier HPC system, which for 184 nodes or more is a maximum of 12 hours. With the learning rate as high as it could comfortably be set to avoid exploding gradients, this limitation on walltime strongly limited the number of steps that could be taken and therefore the final loss that could be obtained. 

\section{Model Merging}

Model merging, also known as parameter averaging, has emerged as a powerful technique for combining capabilities of multiple expert models into a single language model \citep{yadavWhatMattersModel2024, dassanaike-pereraCutsStitchesDoes2023}. While our CPT+SFT procedure significantly improved the model's astronomical knowledge recall in few-shot prompts, we observed that performance in conversational \qa scenarios such as multi-turn conversations and instructions regarding the output style still fell slightly short of optimal. This challenge likely stems from the fact that the ``instruct'' version of Llama-3.1-8B provided by Meta underwent substantially more extensive supervised fine-tuning than what we could achieve as an academic group. We found that merging our specialized model with Meta's instruct model significantly improved these conversational capabilities. 

To create the final version of \model, we employed \textsc{mergekit} \citep{goddardArceesMergeKitToolkit2024}, using the DARE-TIES method to combine our SFT-trained model described in \S\ref{sec:sft} with Meta-Llama-3.1-8B-Instruct \citep{dubeyLlama3Herd2024b}. The merge was performed at full density, BF16 precision, and with the weight parameters set to 0.75 and 0.25 for \model-SFT and Meta-Llama-3.1-8B-Instruct, respectively. 

The resulting merged model exhibits enhanced instruction-following capabilities and improved performance on the AstroMLab-1 multiple-choice question benchmark in both few-shot and structured output scenarios. To determine whether these improvements stemmed from enhanced instruction-following rather than additional astronomical knowledge, we conducted a control experiment. We fine-tuned a separate version of the CPT+SFT model on unrelated multiple-choice questions using identical output formatting. This control model achieved near-identical scores on the AstroMLab-1 benchmark in the structured output scenario without any merging, suggesting that the process of merging in a small fraction of the Meta-Llama-3.1-8B-Instruct model weights transferred general question-answering capabilities rather than domain-specific knowledge.

\begin{figure*}
    \centering
    \includegraphics[width=\linewidth]{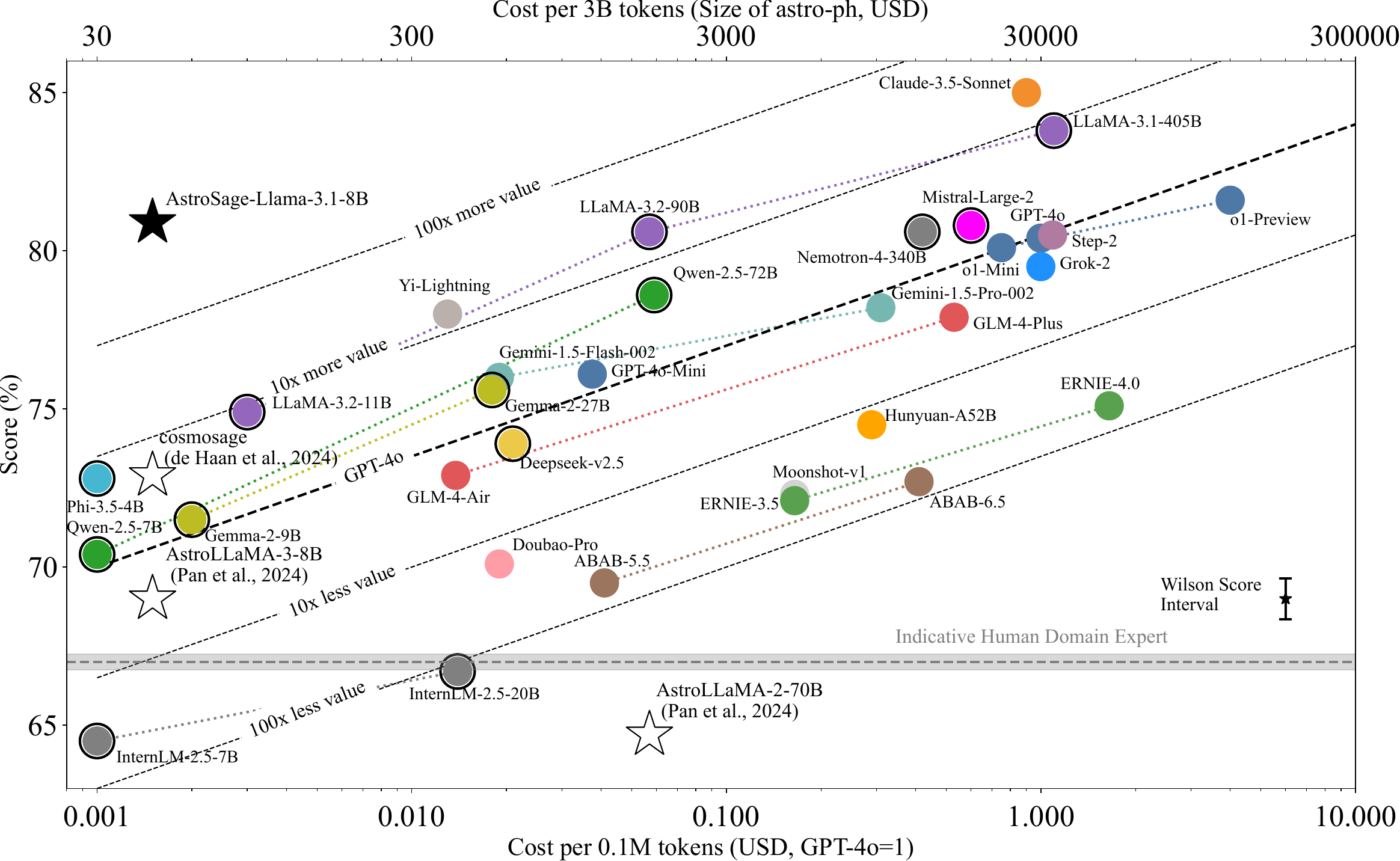}	
    \caption{Performance comparison on the AstroMLab 1 benchmark, which contains 4,425 high-quality, human-verified multiple-choice questions across astronomy, astrophysics, cosmology, and instrumentation. \refresp{The gray shaded bar spans the performance of the two human domain experts.} We present updated results as of November 2024, incorporating both cutting-edge proprietary and open-weight models. \refresp{Open-weight models are circled.} \model outperforms all other models in the 8-billion parameter class and achieves performance comparable to OpenAI's latest models, including GPT-4o, while Claude-3.5-Sonnet maintains the highest performance overall. The diagonal dashed lines represent cost-efficiency trade-offs as determined in AstroMLab 1 (see text for details). The Wilson Score interval shows the typical uncertainty in the score due to the finite number of questions. Star symbols indicate all published specialized LLMs for astronomy to our knowledge. Previously, these specialized models often failed to outperform their baseline models in astronomical recall due to various training limitations. \model represents a significant advancement in specialized astronomical LLMs, demonstrating that extensive data curation, massive continued pre-training and supervised fine-tuning, and model merging techniques can substantially improve performance on specific astronomical tasks. This result highlights the effectiveness of domain specialization even in relatively smaller models.
    }
    \label{fig:benchmark_results}
\end{figure*}

To recap, we began with Meta-Llama-3.1-8B as our base model, then performed CPT on a large corpus of astronomy literature to instill domain knowledge. This was followed by SFT using carefully curated instruction-response pairs to improve task performance and instruction following. Finally, we merged the resulting model with Meta-Llama-3.1-8B-Instruct to enhance general instruction-following capabilities while preserving the astronomical expertise, resulting in our final model which we are releasing as \model.

\section{Evaluation}
\label{sec:evaluation}

\refresp{We have performed three types of quantitative evaluation. The first is an evaluation of \model on a multiple-choice question benchmark, the second on several general-purpose benchmarks, and the third direct preference testing against a similar model that was not specialized in astronomy.}

\subsection{Multiple-Choice Question Benchmark}

\begin{figure*}
    \centering
    \includegraphics[width=\linewidth]{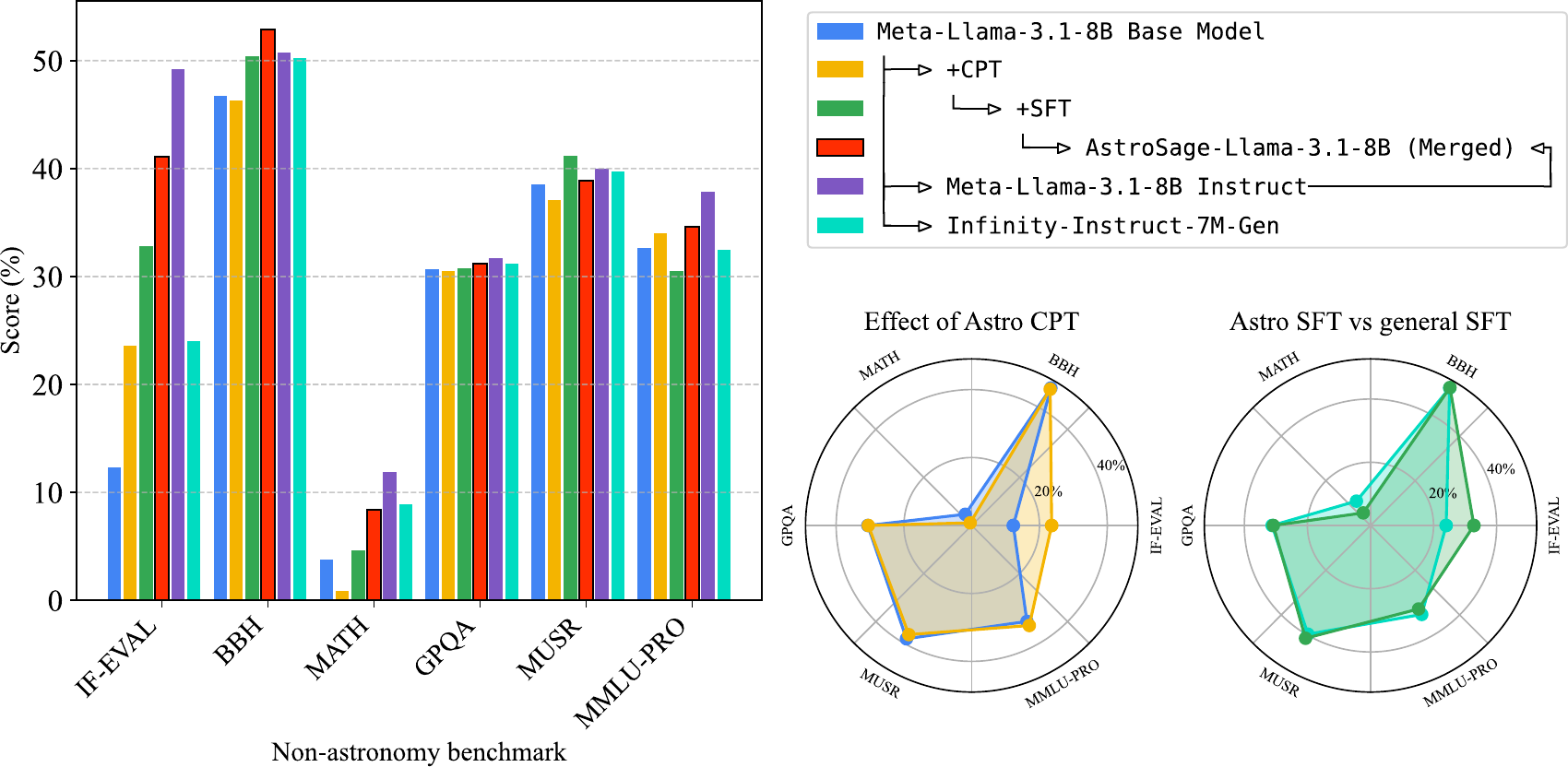}	
    \caption{Performance comparison across general language model benchmarks. \refresp{The bar chart shows a comparison of model performances on standard benchmarks.} The \refresp{left} radar chart visualizes the effect of continued pretraining. The right radar chart compares our model post-SFT against BAAI/Infinity-Instruct-7M-Gen-LLaMA3\_1-8B, which was trained on the same base model and SFT data but without our astronomy-specific data. Despite optimization for astronomy tasks, the merged model we are releasing as \model maintains strong general capabilities in reasoning, mathematics, and coding, demonstrating that domain specialization did not come at the cost of other abilities.}
    \label{fig:general_benchmark}
\end{figure*}

To evaluate \model's performance, we employed the multiple-choice question benchmark from the first paper in this series \citep[AstroMLab 1]{tingAstroMLab1Who2024}. This benchmark consists of diverse astronomy-related questions generated from selected Annual Review of Astronomy and Astrophysics (ARAA) papers and remains, to our knowledge, the only comprehensive astronomy-specific benchmarking effort available. We refer interested readers to the original paper for detailed benchmark specifications.

Importantly, we deliberately excluded the ARAA papers from \model's training dataset. This strategic exclusion enables us to evaluate the model's broader understanding of astronomical concepts rather than its ability to recall specific information from the source materials. This approach helps ensure that the benchmark scores reflect \model's genuine comprehension of astronomy rather than mere memorization of the content used to create the questions.

Our choice to primarily evaluate \model with a knowledge-based benchmark was motivated by two key factors. First, this benchmark represents the only extensively tested and human-vetted dataset available for astronomical knowledge assessment. Second, while astronomical knowledge recall represents just one aspect of LLM capabilities, it serves as a critical foundation for more advanced applications such as scientific agents. The primary goal is to demonstrate that proper fine-tuning of a relatively small model can significantly improve performance on a specific task---an achievement not previously demonstrated in astronomy.

The performance score is calculated as the fraction of correctly answered multiple-choice questions in the benchmarking dataset. The resulting scores are shown in Figure~\ref{fig:benchmark_results}, where round symbols represent scores for cutting-edge proprietary and open-weight models. The open-weight models are also marked with an outer circle. The x-axis displays the cost per $10^5$~tokens, a metric chosen based on practical applications: in the first (and to our knowledge, only) implementation of astronomical agents \citep{sun2024}, analyzing a celestial source's spectral energy distribution from James Webb Space Telescope data requires approximately $10^5$~tokens. The top x-axis shows costs scaled to 3B ($3\times10^9$) tokens, roughly equivalent to the entire astro-ph section of the arXiv. For proprietary models, we use current token costs (averaging input and output costs where they differ), while open-weight model costs are estimated based on typical pricing of \refresp{commercial GPU platforms.}

Specialized astronomical LLMs are denoted by star symbols, except for the first AstroLLaMA model \citep{nguyenAstroLLaMASpecializedFoundation2023}, whose score falls below the plot's lower limit. The bottom right panel shows the typical uncertainty (calculated using the Wilson score interval), demonstrating that our dataset of 4,425~multiple-choice questions provides sufficiently small sampling noise to establish robust performance differences. We have updated all scores using the latest model versions following the methodology from \citet[AstroMLab 1]{tingAstroMLab1Who2024}.

The diagonal dashed lines represent a universal cost-efficiency trade-off observed across major model series (e.g. Llama, GPT, GLM) that simultaneously released models at multiple sizes. We consistently observe a 3.5-point improvement in performance for every 10-fold increase in cost across model families. Each dashed line represents this equivalent trade-off, offset by 3.5 percentage points (equivalent to a 10-fold gain in cost-effectiveness). Despite similar performance on general benchmarks, cutting-edge models can differ by up to 1000-fold in cost-effectiveness on astronomical tasks, highlighting the importance of specialized astronomical benchmarks for evaluating performance on niche technical domains.

To establish a human performance baseline, two domain experts from our team independently completed a random subset of benchmark questions under controlled conditions. \refresp{The experts answered on the order of one hundred questions each, taking around 30 seconds per question. No external references, web searches, or language model assistance were used. Both experts achieved remarkably consistent scores of approximately 68\%, which we designate as the ``Indicative Human Domain Expert'' score.} The fact that most evaluated LLMs significantly surpassed this baseline demonstrates both the benchmark's comprehensive scope and difficulty, while highlighting the remarkable capabilities of current LLMs in capturing and applying complex astronomical knowledge.

As previously noted in \citep[AstroMLab 2]{pan_astromlab_2024}, existing specialized astronomical LLMs (shown as open stars in Figure~\ref{fig:benchmark_results}) fail to outperform baseline models of comparable parameter size. In many cases, suboptimal specialization techniques actually led to performance degradation. In contrast, \model, despite its modest size of 8 billion parameters, achieved an accuracy of \score on this benchmark—comparable to OpenAI's latest flagship models (GPT-4o: 80.4\%) and the best 90B-parameter open-weight Meta-Llama models (80.6\%). This performance is particularly notable because \model achieves these results at approximately one-thousandth the inference cost of proprietary models and one-hundredth the cost of open-weight models. Furthermore, it demonstrates an 8-point improvement over its baseline model, Meta-Llama-3.1-8B (72.9\%). To our knowledge, this represents the first demonstration of a specialized astronomical LLM achieving objectively verified improvements through model fine-tuning.

\subsection{General-Purpose Benchmarks}

To ensure our domain specialization \refresp{did not} compromise general capabilities, we evaluated \model across a comprehensive suite of standard language model benchmarks. These include IF-EVAL \citep{zhou_instruction-following_2023} (instruction following), BBH \citep{suzgun_challenging_2022} (binary hypothesis testing), MATH \citep{hendrycks_measuring_2021} (mathematical reasoning), GPQA \citep{rein_gpqa_2023}(graduate-level science questions), MUSR \citep{sprague_musr_2023} (real-world decision-making scenarios), and MMLU-PRO \citep{wang_mmlu-pro_2024} (an expanded version of MMLU with more challenging reasoning questions). As shown in Figure~\ref{fig:general_benchmark}, our CPT+SFT model (green, initialized from the Llama-3.1 base model) initially performed below the Llama-3.1 instruct model (purple) on five out of the six non-astronomy benchmarks. This was expected, given that Meta's proprietary SFT dataset for their instruct model likely far exceeds what is feasible for an academic research group to reproduce. The merging procedure, pulling in only 25\% of its weight from Meta-Llama-3.1-8B-Instruct, allowed us to recover much of this performance deficit. 

Crucially, this performance recovery through model merging did not compromise \model's astronomical expertise—it maintained its 8-point improvement (representing more than 100-fold increase in cost-effectiveness) on astronomical \qa tasks while largely preserving capabilities across most general benchmarks. The only notable performance decrease occurred in IF-EVAL, which tests instruction following. This limited decline is unsurprising, as instruction following remains one of the more brittle capabilities in language models and likely heavily depends on the proprietary training data used in Meta's instruct model. In fact, when comparing \model to BAAI/Infinity-Instruct-7M-Gen-LLaMA3 1-8B, the latter shows an even more severe performance deficit, highlighting how our refined training strategy and expanded SFT dataset represent crucial improvements. Ultimately, our model merging approach successfully preserved most general capabilities without sacrificing the gained astronomical expertise. This balance is essential, as it enables \model to engage in natural conversations and assist with broader tasks while excelling in astronomy-specific applications.

\subsection{Human Blind Rankings}

\refresp{Simultaneously with \citet{de_haan_cosmosage_2025}, we performed a blinded preference ranking of \model against Meta-Llama-3.1-8B-Instruct. Fifteen questions were written about diverse areas of cosmology, spanning layperson to professional-level difficulty. Independent evaluators compared responses from \model and Meta-Llama-3.1-8B-Instruct, with both models receiving identical system prompts tailored to the cosmology context.}

\refresp{Responses were presented in randomized order to the evaluators, who rated their quality without knowledge of the source. The evaluation involved three evaluators, who preferred the \model answers in 73\% of cases, reflecting a statistically significant preference for \model.}

\section{Availability}
\label{sec:availability}

To promote reproducibility and advance the field of domain-specific AI assistants, we are making \model freely available under the highly permissive Llama 3.1 Community License. The full model weights can be accessed and downloaded from our project repository on Hugging Face: \url{https://huggingface.co/AstroMLab/AstroSage-8B} in either PyTorch or safetensors format.

The code used to prepare the datasets and perform the training will be made available upon reasonable request. The synthetically generated \qa pairs are currently supporting research and development of our next-generation model. After the completion of our research trajectory, we anticipate a full release of our synthetic data set. For inquiries regarding data usage, collaboration, or access to specific subsets of our work, interested parties are encouraged to contact the corresponding author.

By making \model widely available, we aim to foster collaboration and innovation in the astronomy community. We encourage researchers to build upon our work and contribute to the ongoing development of specialized AI assistants for scientific domains.

\section{Discussion and Future Work}
\label{sec:discussion}

This work demonstrates the potential of specialized language models in astronomy through a systematic approach to model development and evaluation. While previous efforts like \citet{pan_astromlab_2024} laid important groundwork in domain-specific modeling, the field has faced persistent challenges in achieving performance gains over baseline models, especially in instruction-following tasks. Our multi-stage training process—combining continued pretraining, extensive supervised fine-tuning, and strategic model merging—addresses these challenges, achieving a notable improvement over the baseline model.

These results demonstrate that powerful AI assistants can be developed with relatively small language models when sufficiently specialized. Despite its modest size of 8 billion parameters, \model achieves performance comparable to latest flagship models at a fraction of the cost—approximately one-thousandth of proprietary models and one-hundredth of open-weight models. This remarkably favorable performance-to-parameter ratio suggests even greater potential for improvement through scaling. Given access to the necessary computational resources, we plan to apply our successful CPT/SFT procedure to a 70B-class model to pursue state-of-the-art astronomy-specific performance.

Beyond the performance achievements, our work establishes a more systematic approach to model evaluation in astronomy. Through tailored astronomy-specific benchmarking in, we provide a more rigorous and transparent assessment than previously available. However, significant challenges remain in comprehensive model evaluation. The field currently lacks standardized, astronomy-specific benchmarks capable of assessing understanding across the full spectrum of astronomical tasks, particularly in exact problem-solving capabilities like those tested in ScienceAgentBench \citep{chen2024}. This limitation restricts our ability to validate comparisons in more direct scientific agent contexts.

The constraints of an 8B-parameter model also become apparent in certain scenarios. While \model demonstrates impressive performance in subjective testing, the AstroMLab-1 benchmark, and general benchmarks, it encounters natural limitations in memory capacity and reasoning depth. Particularly challenging are questions requiring complex multi-step reasoning or sophisticated calculations, where larger general-purpose models still maintain an advantage.

To address these limitations, our future work will pursue several complementary directions. While scaling up model size remains a primary goal, we will also focus on developing more specialized benchmarking tools and exploring retrieval-augmented generation for improved knowledge access. Additional initiatives include creating multilingual astronomy assistants, implementing mechanisms for real-time knowledge updates, and providing public inference capabilities.

The broader implications of this work extend well beyond its immediate achievements. \model serves as a compelling proof of concept for highly specialized, smaller-scale language models in astronomy. Our approach of extensive data curation, continued pretraining, and careful supervised fine-tuning demonstrates how domain-specific expertise can be enhanced while preserving general capabilities. As the field progresses toward agentic research assistants capable of autonomous literature review, data analysis, and hypothesis generation, the need for affordable, highly competent domain-specific models will only grow. While challenges remain, \model charts a promising course for developing the next generation of specialized scientific AI assistants, potentially transforming how we approach astronomical research and education.

\section*{Acknowledgements}

This research used resources of the Oak Ridge Leadership Computing Facility (OLCF), which is a DOE Office of Science User Facility at the Oak Ridge National Laboratory supported by the U.S. Department of Energy under Contract No. DE-AC05-00OR22725 and support from Microsoft's Accelerating Foundation Models Research (AFMR) program. TdH was supported by World Premier International Research Center Initiative (WPI), MEXT, Japan. YST is supported by the National Science Foundation under Grant No. 2406729. Work at Argonne National Lab is supported by UChicago Argonne LLC, Operator of Argonne National Laboratory. Argonne, a U.S. Department of Energy Office of Science Laboratory, is operated under contract no. DE-AC02-06CH11357. A special thanks goes out to Cassie Reuter and Joshua Montgomery for acting as independent evaluators.

\makeatletter
\renewcommand\bibsection{\section*{\MakeUppercase{\refname}}}
\makeatother

\bibliography{astrosage}

\end{document}